\documentclass[12pt]{article}
\usepackage{geometry}                
\usepackage{graphicx}
\usepackage{bm}
\usepackage{amssymb,amsmath,amsthm}
\usepackage{mathrsfs} 
\usepackage{epstopdf}
\def\TODAY{27 August 2010}

\title{\bf Any spacetime has a Bianchi type~I spacetime as a limit}
\author{{\bf Bethan Cropp} and {\bf Matt Visser}\\
School of Mathematics, Statistics, and Operations Research, \\
Victoria University of Wellington, \\
Wellington, New Zealand
}
\date{\TODAY;  \LaTeX-ed \today}                                           
\begin{document}
\maketitle
\begin{abstract}
Pick an arbitrary timelike geodesic in an arbitrary spacetime. We demonstrate that there is a particular limiting process, an ``ultra-local limit'', in which the immediate neighborhood of the timelike geodesic can be ``blown up'' to yield a general (typically non-diagonal) Bianchi type~I spacetime.  This process shares some (but definitely not all) of the features of the Penrose limit, whereby the immediate neighborhood of an arbitrary null geodesic is ``blown up'' to yield a $pp$-wave as a limit. 
\\
\\
File: {\sf  \jobname .tex}

\end{abstract}
\clearpage
\def\d{{\mathrm{d}}}
\newcommand{\scri}{\mathscr{I}}
\newcommand{\sun}{\ensuremath{\odot}}
\def\J{{\mathscr{J}}}
\def\sech{{\mathrm{sech}}}
\def\T{{\mathcal{T}}}
\def\tr{{\mathrm{tr}}}
\hrule
\tableofcontents

\bigskip
\hrule

\clearpage
\section{Introduction}
We shall demonstrate below that there is a particular limiting process, an ``ultra-local limit'', in which the immediate neighborhood of any arbitrary timelike geodesic of any arbitrary spacetime can be  ``blown up'' to yield a general (typically non-diagonal) Bianchi type~I spacetime.  This ultra-local limiting process is  inspired by the well-known ``Penrose limit''~\cite{Penrose1,Penrose2, pp1, pp2, exact1, exact2}, whereby the immediate neighborhood of an arbitrary null geodesic in an arbitrary spacetime is ``blown up'' to yield a $pp$-wave.  

While the ultra-local limit we construct below and the Penrose limit have some key features in common, there are also some significant differences --- in particular the ``hereditary property'' (see particularly reference~\cite{pp2}) fails for the ultra-local limit, which does significantly constrain its usefulness.  On the other hand, given the enormous amount of attention that has been paid to Bianchi type~I spacetimes for purely cosmological reasons, (for a selection of classic and recent sources see~\cite{exact1, exact2, Misner, Collins, Ryan, Uggla, Calogero1, Calogero2, Uzan1, Uzan2}), there is a lot of material  available to help us understand the limiting Bianchi type~I spacetimes. One particular caveat that should be borne in mind is that in a cosmological setting almost all authors rapidly restrict attention to Bianchi type~I spacetimes with a diagonal metric --- this is not appropriate when dealing with the ultra-local limit, and typically we must retain full generality and consider non-diagonal Bianchi type~I spacetimes.

Note further that a closely related limit has sometimes been used in developing mini-superspace models for canonical quantum gravity~\cite{mini1,mini2}, while a related approximation underlies the so-called ``velocity-dominated'' cosmologies~\cite{velocity-dominated, gowdy} and the BKL singularities~\cite{BKL}.
The focus herein is rather different, and we shall be more interested in the classical properties of the Bianchi type~I spacetimes that emerge from this ultra-local limit.

\section{Ultra-local limit}

To define the ultra-local limit pick an arbitrary timelike geodesic $\gamma$ and, in some timelike tube surrounding the chosen curve, choose a specific set of adapted coordinates coordinates $(t,x^i)$ such that:
\begin{itemize}
\item The curve in question lies at $x^i = 0$.
\item At  $x^i = 0$ the coordinate $t$ is just proper time along the curve.
\item The coordinate system is synchronous.
\end{itemize}
This coordinate choice is enough to set
\begin{equation}
\d s^2 = - c^2 \d t^2 + g_{ij}(t,x) \; \d x^i \d x^j,
\end{equation}
where we have explicitly introduced the speed of light. 
Now implement the ultra-local limit in stages:
\begin{enumerate}
\item 
Make the coordinate transformation $x^i \to \epsilon x^i$. Then
 \begin{equation}
 \d s^2 \to - c^2 \d t^2 + \epsilon^2 g_{ij}(t,\epsilon x) \; \d x^i \d x^j.
 \end{equation}
\item
Change the speed of light $c \to \epsilon c$. So  the net effect is 
\begin{equation}
\d s^2 \to - \epsilon^2 c^2 \d t^2 + \epsilon^2 g_{ij}(t,\epsilon x) \; \d x^i \d x^j.
\end{equation}
\item
Perform a conformal transformation $  \d s^2 \to \epsilon^{-2} \d s^2$. So overall
\begin{equation}
\d s^2 \to - c^2 \d t^2 +  g_{ij}(t,\epsilon x) \; \d x^i \d x^j.
\end{equation}
\item 
Take the limit $\epsilon\to 0$. Then the net effect is 
\begin{equation}
\d s^2 \to - c^2 \d t^2 +  g_{ij}(t) \; \d x^i \d x^j.
\end{equation}
\end{enumerate}
This definition is clearly ``ultra-local'' --- one has effectively set $c\to 0$ so that nothing propagates, so that  every point in ``space'' becomes disconnected from every other point; and each individual point in space is then ``blown up'' by the conformal transformation to yield an independent universe. This closely parallels the Penrose limit construction~\cite{Penrose1, Penrose2, pp1, pp2}, in fact one can argue that it is as close as one can possibly get to a ``timelike Penrose limit'' --- though one should certainly not expect all of the features of the usual ``lightlike Penrose limit'' to survive in this timelike case. 

Note that the the 3-dimensional spatial slices are examples of the simplest (type~I) element of Bianchi's classification of translation invariant 3-dimensional Riemannian geometries~\cite{Bianchi}, so the output from this ultra-local limit is is exactly the definition of a \emph{general} Bianchi type~I spacetime~\cite{exact1, exact2, Misner, Collins, Ryan, Uggla, Calogero1, Calogero2, Uzan1, Uzan2} --- the \emph{general} definition before you make any assumptions about diagonalizability of the spatial metric $g_{ij}(t)$. Note further that if one starts with a \emph{congruence} of timelike geodesics in the original spacetime, then the output of this ultra-local limit is a \emph{collection} of general Bianchi type~I spacetimes, one Bianchi spacetime being attached to each timelike geodesic. 
\section{General Bianchi type~I spacetime}
Now that the speed of light $c$ has done its job, let us quietly adopt units where $c=1$, so that we are considering
\begin{equation}
\d s^2 = -\d t^2 + g_{ij}(t) \; \d x^i \d x^j.
\end{equation}
Geometrically this is a rather simple spacetime. Define the extrinsic curvature
\begin{equation}
K_{ij} = -{1\over2} {\d g_{ij}\over\d t},
\end{equation}
then in particular
\begin{equation}
\tr(K) = - {1\over\sqrt{g_3}}  \; {\d \sqrt{g_3}\over\d t},
\end{equation}
where the notion of ``trace'' implicitly involves appropriate factors of the 3-metric and its inverse.
Specializing the usual ADM decomposition, the key results for the Riemann tensor are 
\begin{equation}
R_{ijkl} =  K_{ik} K_{jl}- K_{il} K_{jk};   
\end{equation}
and 
\begin{equation} 
R_{0i0j} = {\d K_{ij}\over\d t} + K_{im} K^m{}_j = {\d K_{ij}\over\d t} + (K^2)_{ij};
\end{equation}
with all other components being zero. 
For an explicit computation, see (and appropriately modify) the discussion of the ADM decomposition in Misner, Thorne, and Wheeler~\cite{MTW}. 
Note carefully that this ultra-local limit does \emph{not} share the hereditary properties of the ordinary Penrose limit. In particular, the ultra-local limit of a Ricci flat spacetime is not necessarily Ricci flat --- from the ADM view this is with hindsight obvious since during the limiting process the extrinsic curvatures and the intrinsic (spatial) curvatures scale in a different manner.  

\subsection{Ricci tensor and scalar}

A brief computation yields
\begin{eqnarray}
R_{00} 
&=& {\d \tr(K)\over\d t } - \tr(K^2),
\end{eqnarray}
while
\begin{eqnarray}
R_{ij} 
& =& - {\d K_{ij}\over\d t} - 2 K_{im} K^m{}_j + \tr(K) K_{ij}.
\end{eqnarray}
Therefore
\begin{eqnarray}
R 
&=& -2{\d \tr(K)\over\d t}+   [\tr(K)]^2  +  \tr(K^2).
\end{eqnarray}

\subsection{Einstein tensor}

The Einstein tensor can also be evaluated from the above, or can be read off from Misner, Thorne, and Wheeler~\cite{MTW}. See (21.162 a,b,c) p 552. The $time$-$time$ component is easy
\begin{eqnarray}
G_{00} 
= 
{1\over2} \left\{ [\tr(K)]^2 - \tr(K^2) \right\}.
\end{eqnarray}
In contrast the $space$-$space$ components are relatively messy
\begin{eqnarray}
G_{ij} 
= - {\d\over\d t} \left[ K_{ij} - g_{ij} \tr(K) \right] + 3 \tr(K) K_{ij} - 2 (K^2)_{ij}- {1\over2} g_{ij} \left\{    [\tr(K)]^2  +  \tr(K^2)  \right\}.
\nonumber\\
\end{eqnarray}

\subsection{Mixed components}

It is often advantageous to work with  \emph{mixed} components $R^a{}_b$. Working from the above, (or suitably adapting results from the key article on BKL spacetimes~\cite{BKL}), one has:
\begin{equation}
R^0{}_0 =  - {\d \tr(K)\over\d t } + \tr(K^2);
\end{equation}
\begin{equation}
R^i{}_0 = 0;
\end{equation}
\begin{equation}
R^0{}_i = 0;
\end{equation}
\begin{equation}
R^i{}_j = -{1\over \sqrt{g_3}} {\d\over\d t} \left[ \sqrt{g_3} \; K^i{}_j \right].
\end{equation}
Also note that this implies (summing only over the space indices)
\begin{equation}
R^i{}_i = {1\over \sqrt{g_3}} {\d^2\over\d t^2} \sqrt{g_3}. 
\label{E:xxx}
\end{equation}
For the Einstein tensor the interesting piece is
\begin{equation}
G^i{}_j = -{1\over \sqrt{g_3}} {\d\over\d t} \left[ \sqrt{g_3} \; K^i{}_j \right] 
+ \delta^i{}_j \left\{  {\d \tr(K)\over\d t}   - {[\tr(K)]^2\over2}  -  {\tr(K^2)\over2} \right\}.
\end{equation}

\subsection{Unimodular decomposition}
It is very useful to decompose
\begin{equation}
g_{ij}(t) = a(t)^2 \; \hat g_{ij}(t),
\end{equation}
where $\det(\hat g_{ij} ) = 1$. Here $a(t)$ can be viewed as an overall scale factor (similar to that occurring in FLRW cosmologies), while $\hat g_{ij}$ describes the ``shape'' of space.
Then
\begin{equation}
{\d g_{ij}\over\d t}  = 2 a \dot a \; \hat g_{ij} + a^2\; {\d{\hat g_{ij}}\over\d t},
\end{equation}
and so
\begin{equation}
K_{ij} = - a \dot a \; \hat g_{ij} + a^2 \; \hat K_{ij},
\end{equation}
with $\hat g^{ij} \,\hat K_{ij} = 0$ because of the unit determinant condition. 
Now define $\hat K^i{}_j = \hat g^{ik} \, \hat K_{kj}$ so that
\begin{equation}
K^i{}_{j} = -  {\dot a\over a} \; \delta^i{}_{j} +  \hat K^i{}_{j}.
\end{equation}
Thus
\begin{equation}
\tr(K) = - 3\;  {\dot a\over a}; 
\qquad
\tr(K^2) = 3 \; {\dot a^2\over a^2} + \tr( \hat K^2 ).
\end{equation}
Then
\begin{equation}
R^0{}_0 =  3 {\ddot a\over a } + \tr(\hat K^2);
\end{equation}
and
\begin{equation}
R^i{}_j = - {\d\over\d t} \left[\hat K^i{}_j \right] - 3 {\dot a\over a} \hat K^i{}_j 
+ \left\{{\ddot a\over a} + 2 {\dot a^2\over a^2}  \right\} \delta^i{}_j.
\end{equation}
Summing over the space components only we have (in agreement with the previous subsection, see equation (\ref{E:xxx}))
\begin{equation}
R^i{}_i =
3 \left\{{\ddot a\over a} + 2 {\dot a^2\over a^2}  \right\}  = {1\over a^3} {\d^2(a^3)\over\d t^2}.
\end{equation}
For the Einstein tensor the interesting piece is
\begin{equation}
G^i{}_j = - {\d\over\d t} \left[\hat K^i{}_j \right] - 3 {\dot a\over a} \hat K^i{}_j 
- {1\over2} \left\{{\ddot a\over a} + 2 {\dot a^2\over a^2}  \right\} \delta^i{}_j.
\end{equation}

\subsection{Summary}

The spacetime curvature of the general Bianchi type~I spacetime can be evaluated as simple algebraic combinations of the $3\times3$ matrices $g_{ij}$, $\dot g_{ij}$, and $\ddot g_{ij}$. If we split the geometry into an overall scale factor $a(t)$ and a shape $\hat g_{ij}$, then the curvature can be evaluated as simple algebraic combination of $a$, $\dot a$, $\ddot a$ and the $3\times3$ matrices $\hat g_{ij}$, $\dot {\hat g}_{ij}$, and $\ddot {\hat g}_{ij}$.

\section{Stress-energy in the ultra-local limit}

What happens to the stress-energy in the ultra-local limit we are interested in? Based on physical intuition one expects that for the energy-momentum flux $T_{0i} \to 0$, but what about $T_{00}$ and $T_{ij}$? 
(Remember that for the light-like Penrose limit $T_{ab}\to T_{uu}$ with all other components vanishing~\cite{Penrose1, Penrose2, pp1, pp2}.) It is convenient to suppress explicit occurrences of $c$ and to write the ultra-local limit in a slightly different form: \\
1) Make the replacement
\begin{equation}
g_{ab}(t,x) \to g^\epsilon_{ab}(t,x) = g_{ab}(t, \epsilon x).
\end{equation}
2) Furthermore, for any generic field $\Psi(t,x)$ make the replacement
\begin{equation}
\Psi(t,x) \to \Psi_\epsilon(t,x) = \Psi(t, \epsilon x).
\end{equation}
3) Then consider the limit as $\epsilon\to0$.

\subsection{Scalar field} 

Consider a scalar field $\phi(t,x)$ with stress-energy tensor
\begin{equation}
T_{ab} = \phi_{,a} \phi_{,b} - {1\over2} g_{ab} \left\{ (\nabla\phi)^2 + V(\phi) \right\}.
\end{equation}
Now take the  ultra-local limit.  Note 
\begin{equation}
T_{00}^{\epsilon} = {1\over2}\left\{ \dot \phi_\epsilon^2 +  g^{ij}_\epsilon \partial_i \phi_\epsilon \partial_j  \phi_\epsilon + V\right\}
= {1\over2}\left\{  \dot \phi^2 + \epsilon^2 g^{ij} \partial_i \phi \partial_j \phi +V \right\} \to  {1\over2} \left\{ \dot \phi^2(t,0) + V \right\},
\end{equation}
whence ultimately
\begin{equation}
T_{00}^{\epsilon}  \to  {1\over2} \left\{ \dot \phi^2(t,0) + V \right\}.
\end{equation}
Similarly
\begin{equation}
T_{0i}^{\epsilon} = {1\over2}\left\{  \dot \phi_\epsilon \partial_i \phi_\epsilon\right\} \to {1\over2}\left\{ \epsilon \dot \phi \partial_i \phi \right\} \to 0.
\end{equation}
Furthermore
\begin{equation}
T_{ij}^{\epsilon} =  \partial_i \phi_\epsilon \partial_j \phi_\epsilon -  {1\over2} g_{ij}^{\epsilon} \left\{  (\nabla\phi)^2 + V \right\} \to 
 \epsilon^2 \partial_i \phi \partial_j \phi -  {1\over2}\left\{ (-\dot\phi^2 + \epsilon^2 (\partial\phi)^2 ) + V \right\} \end{equation}
whence ultimately
\begin{equation}
T_{ij}^{\epsilon} \to  \left\{ \dot \phi^2(t,0) - V \right\} g_{ij}.
\end{equation}
The key point is that the structure of the stress-energy tensor guarantees that in the ultra-local limit 
\begin{equation}
T_{00} \to \rho;
\qquad
T_{0i} \to 0;
\qquad
T_{ij} \to p \; g_{ij}.
\end{equation}
This is a co-moving perfect fluid. Such simple behaviour is common, by no means universal.

\subsection{Electromagnetic field} 

For the electromagnetic field the most natural form of the ultra-local limit is
\begin{equation}
A^a(t,x) \to A_\epsilon^a(t,x) = \left( \phi(t,\epsilon x); \; \vec A(t,\epsilon x) \right),
\end{equation}
in which case 
\begin{equation}
\vec E \to - \dot {\vec A}(t, 0);  \qquad \vec B \to 0.
\end{equation}
So for the stress-energy tensor
\begin{equation}
T_{00} \to  {1\over2} \left\{ g^{ij} \dot A_i \dot A_j\right\};
\end{equation}
\begin{equation}
T_{0i} \to 0;
\end{equation}
\begin{equation}
T_{ij} \to \dot A_i \dot A_j -   {1\over2} \left\{ g^{kl} \dot A_k \dot A_l\right\} g_{ij}
\end{equation}
In terms of the electric field $E$, and the naturally defined inner product $g(E,E)$, it is best to summarize this as:
\begin{equation}
T_{00} \to  {1\over2} \; g(E,E);
\end{equation}
\begin{equation}
T_{0i} \to 0;
\end{equation}
\begin{equation}
T_{ij} \to E_i E_j-   {1\over2} \; g(E,E) g_{ij}.
\end{equation}
Note that in general the $3\times3$ matrix of $space$-$space$ components of the stress-energy will not be diagonal. 

Note further that in reference~\cite{Ryan} a somewhat different limit is taken --- because those authors are interested in different physics. This is done because if one is interpreting the Bianchi type~I spacetime as a cosmology, as an approximation to the large-scale structure of our own physical universe, then for physical reasons one might wish to allow $B\neq0$, while expecting that $E\to0$. Nevertheless, even in this very different context, one key point remains: In general the $3\times3$ matrix of $space$-$space$ components of the stress-energy need not and will not be diagonal.

\subsection{Perfect fluid} 
Consider a generic perfect fluid in a generic spacetime,
\begin{equation}
T^{ab} = (\rho+p) V^a V^b + p g^{ab},
\end{equation}
and ask what the ultra-local limit might be? Working from the geometrical side we have $G^{0i}\to 0$ in the naturally adapted coordinate system. Thus via the Einstein equations it follows that  we must also have $T^{0i} \to 0 $, which in turn implies $V^i \to 0$. Therefore 
\begin{equation}
T_{00} \to \rho;
\qquad
T_{0i} \to 0;
\qquad
T_{ij} \to p \; g_{ij}.
\end{equation}
That is, the generic perfect fluid reduces in the ultra-local limit to a co-moving perfect fluid.

\subsection{Anisotropic fluid}
 Consider a generic anisotropic fluid 
\begin{equation}
T^{ab} = (\rho+p) V^a V^b + p g^{ab} + \Theta^{ab},
\end{equation}
and ask what the ultra-local limit might be? (Here $\Theta$ is the anisotropic stress tensor, which is always 4-traceless and 4-orthogonal to the 4-velocity $V$.) However we choose to define this limit, since we know from the geometrical side that $G^{0i}\to 0$,  we must via the Einstein equations have have $T^{0i} \to 0 $, so that the stress-energy tensor is (3+1) block diagonal.  This in turn implies that we can without loss of generality redefine our variables so that  $V^i \to 0$, while  $V^0\to 1$, and simultaneously choose  $\Theta^{00}\to 0$ along with $\Theta^{0i}\to 0$. Therefore 
\begin{equation}
T_{00} \to \rho;
\qquad
T_{0i} \to 0;
\qquad
T_{ij} \to \pi_{ij} = p g_{ij} + \Theta_{ij}.
\end{equation}

\subsection{Stress-energy conservation}
Consider the covariant conservation law
\begin{equation}
\nabla_a T^{ab}=0 \implies {1\over\sqrt{-g_4}} \partial_a ( \sqrt{-g_4} \; T^{ab}) + \Gamma^b{}_{cd} T^{cd} = 0.
\end{equation}
The only nontrivial component is
\begin{equation}
 {1\over\sqrt{g_3}} \partial_t ( \sqrt{g_3} \; T^{00}) + \Gamma^0{}_{cd} T^{cd}  = 0,
\end{equation} 
which implies 
\begin{equation}
 {1\over\sqrt{g_3}} \partial_t ( \sqrt{g_3} \rho) + {1\over2} {\d g_{ij}\over\d t} \pi^{ij}  = 0,
\end{equation} 
that is
\begin{equation}
 {1\over\sqrt{g_3}} \partial_t ( \sqrt{g_3} \rho) = -{1\over2} {\d g_{ij}\over\d t} \pi^{ij}.
\end{equation} 
To see the connection with more usual FLRW spacetime note $g_{ij} = a^2 \; \hat g_{ij}$ and then
\begin{equation}
 {1\over a^3} \; \partial_t ( a^3 \rho) =  
 -{1\over2}  \left(2 a\dot a \,\hat g_{ij} - a^2 \,{\d\hat g_{ij}\over\d t} \right) \pi^{ij},
\end{equation} 
which we can write as
\begin{equation}
\partial_t ( a^3 \rho) =   -a^2 \dot a g_{ij} \pi^{ij}  -    {1\over2} a^5 {\d\hat g_{ij}\over\d t} \pi^{ij}.
\end{equation} 
That is,  (defining $\pi_{ij} = a^2 \, \hat\pi_{ij}$, so that $\pi^i{}_j = \hat \pi^i{}_j$ and $\pi^{ij} = a^{-2} \, \hat\pi^{ij}$),
\begin{equation}
{\d( \rho \, a^3)\over \d t}  =     - 3 a^2 \dot a p  -    {1\over2} \; a^3 \; {\d\hat g_{ij}\over\d t} \; \{ \hat \pi^{ij} - p \hat g^{ij}\}.
\end{equation}
Finally, (defining $\Theta_{ij} = a^2 \, \hat\Theta_{ij}$, so that $\Theta^i{}_j = \hat \Theta^i{}_j$ and $\Theta^{ij} = a^{-2} \, \hat\Theta^{ij}$),
\begin{equation}
{\d( \rho \, a^3)\over \d t}  =   - p {\d (a^3) \over\d t}   -    {1\over2} \; a^3 \; {\d\hat g_{ij}\over\d t} \; \hat \Theta^{ij}.
\end{equation}
The first 2 terms are the usual energy conservation law. The last term involves the trace-free anisotropic part of the (spatial) stress tensor, together with the trace-free contribution from $\d\hat g_{ij}/\d t$.  We can also write this as
\begin{equation}
{\d\rho\over \d t}  =   - 3(\rho+p) {\d a \over\d t}  + \hat K_{ij}\; \hat \Theta^{ij}.
\end{equation}

\subsection{Summary} 
In general the message is that in the ultra-local limit the stress-energy satisfies
\begin{equation}
T_{00} \to \rho;
\qquad
T_{0i} \to 0;
\qquad
T_{ij} \to \pi_{ij};
\end{equation}
and that one cannot say anything more than this without specifying the particular form of the matter content. Furthermore in this ultra-local limit there is a natural generalization of the energy conservation law normally applied to FLRW spacetimes, with an extra term coming from the interplay between anisotropies in the stress tensor and changes in the shape (not volume) of the spatial slices.

\section{Einstein equations}

In the ultra-local limit the Einstein equations, $G_{ab}  = R_{ab} - {1\over2} R g_{ab} = 8\pi G_N T_{ab}$ reduce to three significant pieces of information --- arising from the the $00$ component, the spatial trace, and the anisotropic part of the spatial trace. It is most convenient to rewrite the Einstein equations as $R_{ab} = 8\pi G_N \{ T_{ab} - {1\over2} T g_{ab}\}$ and then adopt units so that $G_N\to 1$ in which case
\begin{equation}
R_{00} = 4\pi (\rho+3p);
\qquad
R^i{}_i =  12\pi(\rho-p);
\qquad
R_{ij} - {1\over 3} R^k{}_k g_{ij} =8\pi \, \Theta_{ij}. 
\end{equation}
The first of these equations implies
\begin{equation}
{\ddot a\over a} =  {4\pi\over3} (\rho+3p) - {1\over3} \tr( \hat K^2 ).
\end{equation}
That is, as compared to spatially flat FLRW spacetimes, changes in the shape of the spatial slices can now contribute to deceleration.
The second equation yields
\begin{equation}
{\ddot a\over a} + 2 \, {\dot a^2\over a^2} = 4\pi(\rho-p).
\end{equation}
So this combination of terms is what one might expect anyway for any spatially flat FLRW cosmology. 
Finally
\begin{equation}
- {\d\over\d t} \left[\hat K^i{}_j \right] - 3 \, {\dot a\over a} \, \hat K^i{}_j  = 8\pi \, \Theta^i{}_j. 
\end{equation}
That is, evolution of the \emph{shape} of the spatial slices is driven by anisotropies in the stress tensor. 
This is the simplest and most straightforward decoupling of the Einstein equations we have managed to find.

\section{Conditions for a [block]-diagonal metric}
\def\diag{{\mathrm{diag}}}

A key issue in taking this ultra-local limit (and in Bianchi type~I cosmologies generally) is the question of when it is possible (or desirable) to restrict attention to diagonal spatial metrics (or more generally block diagonal metrics). Fortunately, within the context of Bianchi type~I spacetimes it is possible to prove the following qualitative result:
\begin{equation}
 \hbox{([block] diagonal stress-energy)} \Leftrightarrow \hbox{([block] diagonal metric)}. 
\end{equation}
\paragraph{Proof $(\Leftarrow$):} A [block]-diagonal metric $g_{ij}(t)$ implies $\dot g_{ij}(t)$ is [block]-diagonal, implies $K_{ij}(t)$ is [block]-diagonal, implies $\dot K_{ij}(t)$ is [block]-diagonal, implies $R_{ij}(t)$ is [block]-diagonal, implies $G_{ij}(t)$ is [block]-diagonal implies $\pi_{ij}(t)$ is [block]-diagonal, implies $\pi^i{}_j(t)$ is [block]-diagonal. (Equivalently, this implies $\Theta_{ij}(t)$ is [block]-diagonal, which implies $\Theta^i{}_j(t)$ is [block]-diagonal.)

\paragraph{Proof $(\Rightarrow)$:}
At time $t_0$ there is no loss of generality in picking coordinates such that both $[g_0]_{ij} = \delta_{ij}$, and such that $[\dot g_0]_{ij} $ is diagonal. Then without loss of generality there is a coordinate system such that $[K_0]_{ij}$ and $[K_0]^i{}_j$ are diagonal.
Our definition of [block]-diagonal stress-energy will be to assert that in this particular coordinate system $T^i{}_{j}(t)$ is [block]-diagonal, whence $G^i{}_j(t)$ is [block]-diagonal, and so $R^i{}_j(t)$ is [block]-diagonal. 
But then
\begin{equation}
{\d\over\d t} [ \sqrt{g_3} K^i{}_j ] \propto  \{\hbox{[block]-diagonal}(t)\}^i{}_j,
\end{equation}
and so
\begin{equation}
\sqrt{g_3} K^i{}_j  = \sqrt{g_{3,0}} [K_0]^i{}_j  + \{\hbox{[block]-diagonal}(t)\}^i{}_j.
\end{equation}
So by our initial assumption on the diagonal nature of $[K_0]^i{}_j$ we have
\begin{equation}
K^i{}_j  =  \{\hbox{[block]-diagonal}(t)\}^i{}_j.
\end{equation}
That is
\begin{equation}
\forall t:    g^{im} \; \dot g_{mj} =  \{\hbox{[block]-diagonal}(t)\}^i{}_j,
\end{equation}
but then 
\begin{equation}
\forall t:  \dot g_{ij} =   g_{im} \; \{\hbox{[block]-diagonal}(t)\}^m{}_j,
\end{equation}
which can be formally integrated in terms of a time-ordered product
\begin{eqnarray}
[g(t)]_{ij} &=&  [g_0]_{im} \exp\left[ \{\hbox{[block]-diagonal}(t)\}\right]^m{}_j
\\
&=&  \delta_{im} \exp\left[ \{\hbox{[block]-diagonal}(t)\} \right]^m{}_j
\\
&=& \exp\left[ \{\hbox{[block]-diagonal}(t)\} \right]_{ij}
\\
&=& \{\hbox{[block]-diagonal}(t)\}_{ij},
\end{eqnarray}
implying that $[g(t)]_{ij}$ remains [block]-diagonal for all time. 

\paragraph{Summary:} A \emph{necessary} and  \emph{sufficient} condition for a [block]-diagonal metric (in the context of Bianchi type~I spacetimes) is that, in the coordinate system where $[g_0]_{ij}=\delta_{ij}$ and $[K_0]_{ij}$ is diagonal, we have 
\begin{equation}
\pi^i{}_j(t) \hbox{ is [block]-diagonal.}
\end{equation}
The metric and stress tensor then decompose into direct sums:
\begin{equation}
g_{ij} = \oplus_A \, [h_A]_{ij}; \qquad \pi_{ij} = \oplus_A \, \{ p_A [h_A]_{ij} \} ; \qquad  \pi^i{}_j = \oplus_A\, \{ p_A [\delta_A]^i{}_{j}\}.
\end{equation}

\paragraph{Examples:} 
In 3 space dimensions the three canonical examples of this behaviour (for a fully diagonal metric) are:
\begin{enumerate}
\item 
Pefect fluid:
\begin{equation}
\pi_{ij} = p \; g_{ij}.
\end{equation}
That is:
\begin{equation}
\pi^i{}_j =  \left[\begin{array}{ccc}p& 0&0\\ 0 & p  & 0\\ 0&0& p \end{array} \right] = p \; \delta^i{}_j.
\end{equation}
The special sub-case $p=0$ corresponds to dust, and the even more special sub-case $\rho=p=0$ corresponds to vacuum (Kasner solutions) --- with all of these special cases (perfect fluid, dust, vacuum) being very well studied in a cosmological setting.

\item 
Electromagnetic field with the electric field $\vec E$ being an eigenvector of $g$:
\begin{equation}
\pi_{ij} = C_1 g_{ij} + C_2 V_i V_j; \qquad  V^i \propto V_i.
\end{equation}
This is equivalent to:
\begin{equation}
\pi_{ij} =  \left[\begin{array}{ccc}p_1 \; g_{11}& 0&0\\ 0 & p_2 \; g_{22} & 0 \\ 0 & 0 & p_2 \; g_{33} \end{array} \right]; \qquad 
g_{ij} = \left[\begin{array}{ccc}g_{11}& 0 & 0\\ 0 &  g_{22} & 0 \\ 0 & 0 & g_{33} \end{array} \right]. 
\end{equation}
That is:
\begin{equation}
\pi^i{}_j =  \left[\begin{array}{ccc}p_1& 0&0\\ 0 & p_2  & 0\\ 0&0& p_2 \end{array} \right].
\end{equation}

\item General diagonal metric:
\begin{equation}
\pi_{ij} = \diag(p_1,p_2,p_3) \qquad g_{ij} = \diag(g_{11},g_{22},g_{33})
\end{equation}
This is equivalent to:
\begin{equation}
\pi_{ij} =  \left[\begin{array}{ccc}p_1 g\;_{11}& 0&0\\ 0 & p_2 \;g_{22} & 0\\ 0&0& p_3  \;g_{33} \end{array} \right]; \qquad 
g_{ij} = \left[\begin{array}{ccc}g_{11}& 0&0\\ 0 &  g_{22} &0\\0&0&g_{33} \end{array} \right].
\end{equation}
That is:
\begin{equation}
\pi^i{}_j =  \left[\begin{array}{ccc}p_1& 0&0\\ 0 & p_2  & 0\\ 0&0& p_3 \end{array} \right].
\end{equation}

\end{enumerate}
There is also the fully generic non-diagonalizable case corresponding to the stress tensor being a single irreducible $3\times3$ block. (And a partially diagonalizable case corresponding to one irreducible $2\times2$ block plus a singleton $1\times1$ block.) 
These are the only relevant cases for 3-dimensions.

\section{Timelike Raychaudhuri equation}

In a Bianchi type~I spacetime we can simplify the  standard (timelike) Raychaudhuri equation~\cite{Hell, Wald} by noticing that we have a natural timelike congruence defined by $u = \partial_t$, while we also have $u = -(\d t)^\sharp$, or more prosaically $u^a = (1,0,0,0)$. For this timelike congruence the vorticity is zero, $\omega=0$, as is  the acceleration $\frac{\mathsf d u^a}{\mathsf d s}  = u^b \nabla_b u^a =0$. 
Therefore in terms of the expansion tensor $\theta_{ab} = \nabla_{(a} u_{b)}$, (not to be confused with the anisotropic stress $\Theta_{ab}$), we have 
\begin{equation}
R_{ab}u^au^b= -\theta_{ab}\theta^{ab}+\theta^2-\frac{\d \theta}{\d t}.
\end{equation}
Separating out the unit determinant metric, $g_{ij} = a^2 \hat g_{ij}$, and using the standard definitions of shear and expansion we find
\begin{equation}
\theta_{ab}= a \dot a \; \hat g_{ab}+a^2\hat K_{ab}; 
\qquad 
\sigma_{ab}=a^2\hat K_{ab}; 
\qquad 
\sigma^2=\frac{1}{2}\tr(\hat K^2).
\end{equation}
Furthermore $K=-3{\dot a}/{a}$, and hence 
\begin{equation}
\theta = 3\frac{\dot a}{a};
\qquad
\frac{{\d \theta}}{{\d t}} = 3\frac{\ddot a}{a}-3\frac{\dot a^2}{a^2}
\end{equation}
So the Raychaudhuri equation is given by
\begin{equation}
R_{ab} u^a u^b=-\tr(\hat K^2)-3\,\frac{\ddot a}{a}.
\end{equation}
Compare this to both the result for (spatially flat) FLRW cosmology,
\begin{equation}
R_{ab}u^au^b=-3\, \frac{\ddot a}{a},
\end{equation}
to which the Bianchi type~I spacetime reduces when $\hat K_{ab}=0$,
and to the result stated by Collins and Ellis~\cite{Collins} for perfect fluid Bianchi cosmologies,
\begin{equation}
{1\over2}(\rho+3p)=-3\frac{\ddot a}{a}-2\sigma^2.
\end{equation}
Attempting to put the Raychaudhuri scalar into a ``nice'' form, the best possible forms seem to be
\begin{equation}
R_{ab}u^au^b=\frac{1}{2}\,(\rho+3p)+\tr(\hat K^2),
\end{equation}
or alternatively
\begin{equation}
R_{ab}u^au^b=\frac{1}{4}\,(\rho+3p)-\frac{3}{2}\,\frac{\ddot a}{a}.
\end{equation}

\section{Discussion}

Overall the main message to take from this article is that the Bianchi type~I spacetimes are interesting not only for their traditional use in cosmology, where they are the simplest example of homogeneous but anisotropic cosmologies, but also in the much wider context of the  ultra-local limit described in this article. In a very precise manner described above, the Bianchi type~I spacetimes are the result of a limiting process that can be applied to any timelike geodesic in any arbitrary spacetime. This  ultra-local limit, although not sharing all important properties,  seems to be the closest timelike analogue one can construct to the more usual Penrose limit (which is applied to lightlike geodesics to generate $pp$-wave spacetimes). 

In the current article we have first set up the ultra-local limit, and outlined its major features. One key item to note is that in the current context there is typically no good reason to restrict oneself  to \emph{diagonal} Bianchi type~I spacetimes --- it is general (typically non-diagonal) Bianchi type~I spacetimes that are of interest here.  
Modulo this technical issue, the existence of a vast cosmological literature on Bianchi type~I spacetimes represents a significant resource which can be applied to the ultra-local limit of interest in this current article.




\begin{thebibliography}{69}

\bibitem{Penrose1}
 Roger Penrose, 
 ``A remarkable property of plane waves in general relativity", 
 Rev. Mod. Phys. {\bf 37} (1965) 215--220. 
 doi:10.1103/RevModPhys.37.215.

\bibitem{Penrose2}
 Roger Penrose,  
 ``Any spacetime has a plane wave as a limit". 
 In \emph{Differential Geometry and Relativity},  edited by M.~Cahen and M.~Flato. 
 (Kluwer/Riedel, Dordrecht, 1976). Pages 271--275.
 
\bibitem{pp1}
J.~D.~Steele,  ``On generalised $pp$ waves", \\
{\sf http://web.maths.unsw.edu.au/\~{}jds/Papers/gppwaves.pdf}\\
{}[Retrieved 14 April 2010].
 
\bibitem{pp2} 
M.~Blau, Plane Waves and Penrose limits'', \\
{\sf http://www.blau.itp.unibe.ch/lecturesPP.pdf}\\
{}[Retrieved 17 August 2010].

\bibitem{exact1}
 Hans Stephani,  Dietrich Kramer,  Malcolm MacCallum,  Cornelius Hoenselaers,  and Eduard Herlt,
 \emph{Exact Solutions of Einstein's Field Equations}. 
 (Cambridge University Press, Cambridge, 2003). 
 \\
 For $pp$ waves see especially section 24.5. 
 \\
 For the Bianchi classification see especially section 8.2.

\bibitem{exact2}
Jerry Griffiths and Ji\v{r}\'{\i} Podolsk\'y, 
\emph{Exact spacetimes in Einstein's general relativity},
 (Cambridge University Press, Cambridge, 2009). 
 \\
 For $pp$ waves see especially chapter 17. 
 \\
 For the Bianchi classification see especially section 22.1.


\bibitem{Misner}
C.~W.~Misner,
  ``The Isotropy of the universe,''
  Astrophys.\ J.\  {\bf 151} (1968) 431.

\bibitem{Collins}
 C.~B.~Collins and S.~W.~Hawking,
  ``Why is the Universe isotropic?,''
  Astrophys.\ J.\  {\bf 180} (1973) 317.
  
\bibitem{Ryan}
 M.~P.~Ryan, S.~M.~Waller, and L.~C.~Shepley,
 ``Bianchi type electromagnetic cosmology - type~I Hamiltonian'',
Astrophysical Journal, Part 1, {\bf 254}, (1982) 425--436. 
  
 \bibitem{Uggla}
  J.~M.~Heinzle and C.~Uggla,
  ``Dynamics of the spatially homogeneous Bianchi type~I Einstein-Vlasov
  equations,''
  Class.\ Quant.\ Grav.\  {\bf 23} (2006) 3463
  [arXiv:gr-qc/0512031].
  
 \bibitem{Calogero1} 
   S.~Calogero and J.~M.~Heinzle,
  ``Dynamics of Bianchi type~I elastic spacetimes,''
  Class.\ Quant.\ Grav.\  {\bf 24} (2007) 5173
  [arXiv:0706.3823 [gr-qc]].

 \bibitem{Calogero2} 
   S.~Calogero and J.~M.~Heinzle,
  ``Bianchi Cosmologies with Anisotropic Matter: Locally Rotationally Symmetric
  Models,''
  arXiv:0911.0667 [gr-qc].

  
\bibitem{Uzan1}
 T.~S.~Pereira, C.~Pitrou and J.~P.~Uzan,
  ``Theory of cosmological perturbations in an anisotropic universe,''
  JCAP {\bf 0709} (2007) 006
  [arXiv:0707.0736 [astro-ph]].  
   
 \bibitem{Uzan2}
 C.~Pitrou, T.~S.~Pereira and J.~P.~Uzan,
 ``Predictions from an anisotropic inflationary era,''
JCAP {\bf 0804} (2008) 004
[arXiv:0801.3596 [astro-ph]].
  
\bibitem{mini1}
Gerson Francisco,
``The behavior of the gravitational field near the initial singularity'',
General Relativity and Gravitation {\bf18} (1986) 287--308,
doi	10.1007/BF00765888

\bibitem{mini2}
James George Tsacoyeanes,
``Ultralocal limit of the gravitational field coupled to a scalar field'',
Phys. Rev. D {\bf35} (1987)  483--494.

\bibitem{velocity-dominated}
D.~Eardley, E.~Liang and R.~Sachs,
  ``Velocity-dominated singularities in irrotational dust cosmologies,''
  J.\ Math.\ Phys.\  {\bf 13} (1972) 99.

\bibitem{gowdy}
J.~Isenberg and V.~Moncrief,
  ``Asymptotic behavior of the gravitational field and the nature of
  singularities in Gowdy spacetimes,''
  Annals Phys.\  {\bf 199} (1990) 84.

  
\bibitem{BKL}
V.~A.~Belinsky,  I.~M.~Khalatnikov, E.~M.~Lifshitz,
``Oscillatory approach to a singular point in the relativistic cosmology",
Soviet Physics Uspekhi {\bf13} (1971) 745--765. [Usp. Fiz. Nauk {\bf102} (1970)  463--500.]


\bibitem{Bianchi}
L.~Bianchi, 
``Sugli spazii a tre dimensioni che ammettono un gruppo continuo di movimenti'',
 (``On the spaces of three dimensions that admit a continuous group of movements''.) 
 Soc. Ital. Sci. Mem. di Mat. {\bf11} (1898) 267--352.\\
 English translation in:  General Relativity and Gravitation {\bf 33} (2001) 2171--2253.
 
 \bibitem{MTW}
C.~W.~Misner, K.~S.~Thorne, and J.~A.~Wheeler, 
``Gravitation'',
(Freeman, San Francisco, 1973). 
See especially (21.76) p 516, (21.82) p 518.

\bibitem{Hell}
S.W.~Hawking and G.~F.~R.~Ellis, 
\emph{The large scale structure of spacetime}, 
(Cambridge, England, 1972).

\bibitem{Wald}
 R.~M.~Wald, 
 \emph{General Relativity}, 
 (University of Chicago Press, Chicago, 1984).

 
 \end{thebibliography}
\end{document}